\documentclass[aip,amsmath,amssymb,reprint]{revtex4-1}

\usepackage{graphicx}
\usepackage{dcolumn}
\usepackage{bm}

\usepackage[utf8]{inputenc}
\usepackage[T1]{fontenc}
\usepackage{mathptmx}

\usepackage[unicode=true,pdfusetitle, bookmarks=false, breaklinks=false,pdfborder={0 0 0},backref=false,colorlinks=false] {hyperref}

\begin{document}

\preprint{AIP/123-QED}

%
%

\title[How to make a monochromatic electromagnetic knot]{Theoretical proposal for the experimental realisation of a monochromatic electromagnetic knot}

\author{R. P. Cameron}
\email{robert.p.cameron@strath.ac.uk}
\affiliation{SUPA and Department of Physics, University of Strathclyde, Glasgow G4 0NG, UK}

\author{W. L\"{o}ffler}
\affiliation{Huygens-Kamerlingh Onnes Laboratorium, Leiden University, 2333 Leiden, CA, The Netherlands}

\author{K. D. Stephan}
\affiliation{Ingram School of Engineering, Texas State University, San Marcos, Texas USA}

\date{\today}

\begin{abstract}
We propose an antenna designed to generate monochromatic electromagnetic knots and other “unusual electromagnetic disturbances” in the microwave domain. Our antenna is a spherical array of radiating dipolar elements configured to approximate the desired electromagnetic field near its centre. We show numerically that a specific embodiment of the antenna with a radius of $61.2\,\textrm{cm}$ and only $20$ element pairs driven at a frequency of $2.45\,\textrm{GHz}$ can yield linked and torus-knotted electric and magnetic field lines approximating those of an ``electromagnetic tangle'': a monochromatic electromagnetic knot closely related to the well-known Ra{\~n}ada-Hopf type electromagnetic knots but simpler in its construction. The antenna could be used to locally excite plasmas.
\end{abstract}

\maketitle

%
%

\section{\label{Introduction}Introduction}
Electromagnetic knots were first described by Ra{\~n}ada and collaborators in a series of works \cite{Ranada89a, Ranada90a, Ranada95a, Ranada96a} based on earlier studies by Woltjer \cite{Woltjer58a} and Moffatt \cite{Moffatt69a}, making use of the Bateman construction \cite{Bateman16a}. Several properties of these Ra{\~n}ada-Hopf electromagnetic knots have been worked out in theory, including their orbital angular momentum \cite{Irvine08a} and helicity \cite{Ranada92a, Ranada97a, Arrayas12a, Irvine10a, Kedia16a, Trueba96a}, and similar solutions of Maxwell's equations have been found more recently in the form of propagating light beams \cite{Bauer15a, Philbin18a, Karimi19a}. Although analogous structures have been observed experimentally in liquid crystals \cite{Tkalec11a} and in fluid dynamics \cite{Kleckner13a}, no feasible proposal has been given to date for the realisation of a Ra{\~n}ada-Hopf type electromagnetic knot, the closest work being a theoretical proposal involving plasma physics and self-organisation \cite{Arrayas10a, Smiet15a}. A key difficulty is that Ra{\~n}ada-Hopf type electromagnetic knots are \textit{polychromatic}; they have a broad frequency spectrum, which is difficult to realise experimentally.

One of us recently described a collection of “unusual electromagnetic disturbances” \cite{Cameron18a}, which are essentially electromagnetic standing waves \cite{Footnote1} that appear to be well localised in 3D, even in complete vacuum. Included among these are various \textit{monochromatic} electromagnetic knots, closely related to the well-known Ra{\~n}ada-Hopf type electromagnetic knots but simpler in their construction.

In this paper we propose an antenna designed to generate monochromatic electromagnetic knots and other unusual electromagnetic disturbances in the microwave domain. We work in an inertial frame of reference with time $t$ and position vector $\mathbf{r}=x\hat{\mathbf{x}}+y\hat{\mathbf{y}}+z\hat{\mathbf{z}}=s\hat{\mathbf{s}}+z\hat{\mathbf{z}}=r\hat{\mathbf{r}}$, where $x$, $y$ and $z$ are right-handed Cartesian coordinates with associated unit vectors $\hat{\mathbf{x}}$, $\hat{\mathbf{y}}$ and $\hat{\mathbf{z}}$; $s$, $\phi$ and $z$ are cylindrical coordinates with associated unit vectors $\hat{\mathbf{s}}=\hat{\mathbf{s}}(\phi)$, $\hat{\bm{\phi}}=\hat{\bm{\phi}}(\phi)$ and $\hat{\mathbf{z}}$ and $r$, $\theta$ and $\phi$ are spherical coordinates with associated unit vectors $\hat{\mathbf{r}}=\hat{\mathbf{r}}(\theta,\phi)$, $\hat{\bm{\theta}}=\hat{\bm{\theta}}(\theta,\phi)$ and $\hat{\bm{\phi}}=\hat{\bm{\phi}}(\phi)$. SI units are employed throughout, with $\epsilon_0$ being the electric constant, $\mu_0$ being the magnetic constant and $c=1/\sqrt{\epsilon_0\mu_0}$ being the speed of light.

%
%

\section{\label{Unusual electromagnetic disturbances}Unusual electromagnetic disturbances}
In this section we briefly summarise the unusual electromagnetic disturbances described in \cite{Cameron18a}. We work in vacuum.

The simplest disturbances, referred to as ``unusual electromagnetic disturbances of the \textit{first} kind'', can each be thought of as a continuous spherical superposition of plane electromagnetic waves with common amplitude, phase at the origin ($\mathbf{r}=0$), polarisation state and frequency. The resulting electric field $\mathbf{E}=\mathbf{E}(\mathbf{r},t)$ and magnetic field $\mathbf{B}=\mathbf{B}(\mathbf{r},t)$ are
\begin{eqnarray}
\mathbf{E}&=&\Re\Bigg\{E_0\Big[\tilde{\mathtt{A}}'_0\left(\hat{\mathbf{s}}g+\hat{\mathbf{z}}h\right)+\textrm{i}\tilde{\mathtt{B}}'_0\hat{\pmb{\phi}}f\Big]\textrm{e}^{-\textrm{i}\omega_0t}\Bigg\} \label{EUED} \\
\mathbf{B}&=&\Re\Bigg\{\frac{E_0}{c}\Big[\textrm{i}\tilde{\mathtt{A}}'_0\hat{\pmb{\phi}}f-\tilde{\mathtt{B}}'_0\left(\hat{\mathbf{s}}g+\hat{\mathbf{z}}h\right)\Big]\textrm{e}^{-\textrm{i}\omega_0t}\Bigg\} \label{BUED}
\end{eqnarray}
with
\begin{eqnarray}
f&=&\int_0^\pi\textrm{J}_1(k_0\sin\vartheta s)\cos(k_0\cos\vartheta z)\sin\vartheta\textrm{d}\vartheta, \\
g&=&-\int_0^\pi\textrm{J}_1(k_0\sin\vartheta s)\sin(k_0\cos\vartheta z)\sin\vartheta\cos\vartheta\textrm{d}\vartheta \\
h&=&-\int_0^\pi\textrm{J}_0(k_0\sin\vartheta s)\cos(k_0\cos\vartheta z)\sin^2\vartheta\textrm{d}\vartheta,
\end{eqnarray}
where $E_0$ is an electric-field strength, $\tilde{\mathtt{A}}'_0$ and $\tilde{\mathtt{B}}'_0$ are constants that derive from the polarisation state of the waves and $\omega_0=c k_0$ is the angular frequency of the disturbance, $k_0$ being the angular wavenumber. A particular unusual electromagnetic disturbance of the first kind is determined by specifying $\tilde{\mathtt{A}}'_0$ and $\tilde{\mathtt{B}}'_0$. For example: taking $\tilde{\mathtt{A}}'_0=\textrm{i}$ and $\tilde{\mathtt{B}}'_0=0$ corresponds to each wave being linearly polarised along the polar direction and gives an ``electric globule''; taking $\tilde{\mathtt{A}}'_0=0$ and $\tilde{\mathtt{B}}'_0=-\textrm{i}$ corresponds to each wave being linearly polarised along the azimuthal direction and gives an ``electric ring''; taking $\tilde{\mathtt{A}}'_0=1/\sqrt{2}$ and $\tilde{\mathtt{B}}'_0=\textrm{i}\sigma/\sqrt{2}$ with $\sigma=\pm 1$ corresponds to each wave having left- or right-handed circular polarisation and gives an ``electromagnetic tangle'', which can be regarded as a (monochromatic) electromagnetic knot in that it has linked and torus-knotted electric and magnetic field lines.

Unusual electromagnetic disturbances of the first kind can be superposed in various ways to create more exotic structures, referred to as ``unusual electromagnetic disturbances of the \textit{second} kind''.

We consider the generation of an (approximate) electromagnetic tangle in section \ref{A specific embodiment}.

%
%

\section{\label{General antenna design}General antenna design}
In this section we describe our antenna design in general terms.

Our antenna is an array of $2N$ driven elements, each taking the form of a centre-fed dipole of length $L$ driven at angular frequency $\omega_0=2\pi f_0=2\pi/T_0=ck_0=2\pi c/\lambda_0$, where $f_0$ is the frequency, $T_0$ is the period, $k_0$ is the angular wavenumber in vacuum and $\lambda_0$ is the wavelength in vacuum. The elements are located in pairs at $N$ points uniformly distributed on the surface of a sphere \cite{Footnote2} of radius $R$ centred on the origin ($\mathbf{r}=0$). One element of each pair, henceforth referred to as the ``$\mathtt{A}$-type'' element of the pair, is aligned with the polar unit vector $\hat{\bm{\theta}}$ and is so-named because it corresponds to the $\tilde{\mathtt{A}}_0'$ contributions in (\ref{EUED}) and (\ref{BUED}). The other element of the pair, henceforth referred to as the ``$\mathtt{B}$-type'' element of the pair, is aligned with the azimuthal unit vector $\hat{\bm{\phi}}$ and is so-named because it corresponds to the $\tilde{\mathtt{B}}_0'$ contributions in (\ref{EUED}) and (\ref{BUED}). The antenna is embedded in a medium of refractive index $n=c\sqrt{\epsilon\mu}=ck/\omega_0$, where $\epsilon$ is the permittivity, $\mu$ is the permeability and $k$ is the angular wavenumber in the medium. 

Consider the $\mathtt{X}$-type element of the $n$th pair ($\mathtt{X}\in\{\mathtt{A},\mathtt{B}\}$, $n\in\{1,\dots, N\}$). The element is centred on position $\mathbf{r}_n=x_n\hat{\mathbf{x}}+y_n\hat{\mathbf{y}}+z_n\hat{\mathbf{z}}$ ($|\mathbf{r}_n|=R$) and is aligned with the unit vector $\hat{\mathbf{u}}_{\mathtt{X}n}$ ($\hat{\mathbf{u}}_{\mathtt{A}n}=\hat{\bm{\theta}}(\hat{\mathbf{r}}_n)$ and $\hat{\mathbf{u}}_{\mathtt{B}n}=\hat{\bm{\phi}}(\hat{\mathbf{r}}_n)$). The electric current $I_{\mathtt{X}n}=I_{\mathtt{X}n}(l,t)$ in the element is
\begin{eqnarray}
I_{\mathtt{X}n}&=&\Re\Big[I_{0\mathtt{X}n}w\textrm{e}^{-\textrm{i}(\omega_0 t-\varphi_{0\mathtt{X}n})}\Big],
\end{eqnarray}
where $I_{0\mathtt{X}n}$ is the peak electric current, $w=w(l)$ is a function that dictates the spatial distribution of the electric current ($0\le w\le1$), $l$ is a coordinate with origin at $\mathbf{r}_n$ that increases in the direction of $\hat{\mathbf{u}}_{\mathtt{X}n}$ ($-L/2\le l \le L/2$) and $\varphi_{0\mathtt{X}n}$ is a phase constant. Assuming that the elements are half-wave dipoles ($L=\lambda_0/2$), we take
\begin{eqnarray}
w&=&\cos(k_0l).
\end{eqnarray}
Other types of element such as short dipoles can be described using other forms for $w$. The electric field $\mathbf{E}_{\mathtt{X}n}=\mathbf{E}_{\mathtt{X}n}(\mathbf{r},t)$ and magnetic field $\mathbf{B}_{\mathtt{X}n}=\mathbf{B}_{\mathtt{X}n}(\mathbf{r},t)$ generated by the element are calculated by decomposing the element into a line of \textit{infinitesimal} dipolar elements and integrating their contributions \cite{Jackson99a}, giving
\begin{eqnarray}
\mathbf{E}_{\mathtt{X}n}&=&\Re\Bigg\{\int_{-L/2}^{L/2}\frac{\textrm{i}\mu\omega_0 I_{0\mathtt{X}n}w\textrm{e}^{\textrm{i}(k|\mathbf{r}-\mathbf{r}_n-\hat{\mathbf{u}}_{\mathtt{X}n}l|-\omega_0 t+\varphi_{0\mathtt{X}n})}}{4\pi|\mathbf{r}-\mathbf{r}_n-\hat{\mathbf{u}}_{\mathtt{X}n}l|} \nonumber \\
&&\Bigg[\hat{\mathbf{u}}_{\mathtt{X}n}\Bigg(1+\frac{\textrm{i}}{k|\mathbf{r}-\mathbf{r}_n-\hat{\mathbf{u}}_{\mathtt{X}n}l|}-\frac{1}{k^2|\mathbf{r}-\mathbf{r}_n-\hat{\mathbf{u}}_{\mathtt{X}n}l|^2}\Bigg) \nonumber \\
&&-\frac{(\mathbf{r}-\mathbf{r}_n-\hat{\mathbf{u}}_{\mathtt{X}n}l)}{|\mathbf{r}-\mathbf{r}_n-\hat{\mathbf{u}}_{\mathtt{X}n}l|}\frac{\hat{\mathbf{u}}_{\mathtt{X}n}\cdot(\mathbf{r}-\mathbf{r}_n-\hat{\mathbf{u}}_{\mathtt{X}n}l)}{|\mathbf{r}-\mathbf{r}_n-\hat{\mathbf{u}}_{\mathtt{X}n}l|} \nonumber \\
&&\Bigg(1+\frac{3\textrm{i}}{k|\mathbf{r}-\mathbf{r}_n-\hat{\mathbf{u}}_{\mathtt{X}n}l|}-\frac{3}{k^2|\mathbf{r}-\mathbf{r}_n-\hat{\mathbf{u}}_{\mathtt{X}n}l|^2}\Bigg)\Bigg]\textrm{d}l\Bigg\} \nonumber \\
&&\label{EXn} \\
\mathbf{B}_{\mathtt{X}n}&=&\Re\Bigg[\int_{-L/2}^{L/2}\frac{\textrm{i}\mu n\omega_0 I_{0\mathtt{X}n}w\textrm{e}^{\textrm{i}(k|\mathbf{r}-\mathbf{r}_n-\hat{\mathbf{u}}_{\mathtt{X}n}l|-\omega_0 t+\varphi_{0\mathtt{X}n})}}{4\pi c|\mathbf{r}-\mathbf{r}_n-\hat{\mathbf{u}}_{\mathtt{X}n}l|} \nonumber \\
&&\frac{(\mathbf{r}-\mathbf{r}_n-\hat{\mathbf{u}}_{\mathtt{X}n}l)\times\hat{\mathbf{u}}_{\mathtt{X}n}}{|\mathbf{r}-\mathbf{r}_n-\hat{\mathbf{u}}_{\mathtt{X}n}l|}\Bigg(1+\frac{\textrm{i}}{k|\mathbf{r}-\mathbf{r}_n-\hat{\mathbf{u}}_{\mathtt{X}n}l|}\Bigg)\textrm{d}l\Bigg].  \label{BXn}
\end{eqnarray}
The electric field $\mathbf{E}=\mathbf{E}(\mathbf{r},t)$ and magnetic field $\mathbf{B}=\mathbf{B}(\mathbf{r},t)$ generated by the antenna as a whole are superpositions of the fields generated by the individual elements:
\begin{eqnarray}
\mathbf{E}&=&\sum_{\mathtt{X}=\mathtt{A},\mathtt{B}}\sum_{n=1}^N\mathbf{E}_{\mathtt{X}n} \label{Esum} \\
\mathbf{B}&=&\sum_{\mathtt{X}=\mathtt{A},\mathtt{B}}\sum_{n=1}^N\mathbf{B}_{\mathtt{X}n}. \label{Bsum}
\end{eqnarray}
The integrals in (\ref{EXn}) and (\ref{BXn}) can be calculated numerically.

We will now elucidate the conditions under which our antenna functions in the desired manner, specialising to vacuum ($\mu=\mu_0$ and $k=k_0$) for a direct comparison with the results presented in \cite{Cameron18a} and summarised in section \ref{Unusual electromagnetic disturbances}. The antenna is designed to generate an approximation to an unusual electromagnetic disturbance internally, near the origin ($\mathbf{r}=0$). Let us focus, therefore, on a spherical region of radius $r_\textrm{max}$ inside the antenna ($0<r_\textrm{max}<R$), centred on the origin ($0\le|\mathbf{r}|\lesssim r_\textrm{max}$). Assuming that the antenna is many wavelengths in radius ($k_0 R\gg 1$) and that we are in the far-field regime with respect to each of the elements ($k_0 (R-r_\textrm{max})\gg 1$), we neglect near-field terms and thus see (\ref{EXn})-(\ref{Bsum}) reduce to
\begin{eqnarray}
\mathbf{E}&\approx&\Re\Bigg\{\sum_{\mathtt{X}=\mathtt{A},\mathtt{B}}\sum_{n=1}^N\int_{-L/2}^{L/2}\frac{\textrm{i}\mu_0\omega_0 I_{0\mathtt{X}n}w\textrm{e}^{\textrm{i}(k_0|\mathbf{r}-\mathbf{r}_n-\hat{\mathbf{u}}_{\mathtt{X}n}l|-\omega_0 t+\varphi_{0\mathtt{X}n})}}{4\pi|\mathbf{r}-\mathbf{r}_n-\hat{\mathbf{u}}_{\mathtt{X}n}l|} \nonumber \\
&&\Bigg[\hat{\mathbf{u}}_{\mathtt{X}n}-\frac{(\mathbf{r}-\mathbf{r}_n-\hat{\mathbf{u}}_{\mathtt{X}n}l)}{|\mathbf{r}-\mathbf{r}_n-\hat{\mathbf{u}}_{\mathtt{X}n}l|}\frac{\hat{\mathbf{u}}_{\mathtt{X}n}\cdot(\mathbf{r}-\mathbf{r}_n-\hat{\mathbf{u}}_{\mathtt{X}n}l)}{|\mathbf{r}-\mathbf{r}_n-\hat{\mathbf{u}}_{\mathtt{X}n}l|}\Bigg]\textrm{d}l\Bigg\} \label{Ereduction1} \\
\mathbf{B}&\approx&\Re\Bigg[\sum_{\mathtt{X}=\mathtt{A},\mathtt{B}}\sum_{n=1}^N\int_{-L/2}^{L/2}\frac{\textrm{i}\mu_0\omega_0 I_{0\mathtt{X}n}w\textrm{e}^{\textrm{i}(k_0|\mathbf{r}-\mathbf{r}_n-\hat{\mathbf{u}}_{\mathtt{X}n}l|-\omega_0 t+\varphi_{0\mathtt{X}n})}}{4\pi c|\mathbf{r}-\mathbf{r}_n-\hat{\mathbf{u}}_{\mathtt{X}n}l|} \nonumber \\
&&\frac{(\mathbf{r}-\mathbf{r}_n-\hat{\mathbf{u}}_{\mathtt{X}n}l)\times\hat{\mathbf{u}}_{\mathtt{X}n}}{|\mathbf{r}-\mathbf{r}_n-\hat{\mathbf{u}}_{\mathtt{X}n}l|}\textrm{d}l\Bigg]. \label{Breduction1}
\end{eqnarray}
Assuming that the spherical region lies sufficiently well within the antenna ($r_\textrm{max}\ll R$) that the electromagnetic wave produced by each element appears planar, we take 
\begin{eqnarray}
\frac{1}{|\mathbf{r}-\mathbf{r}_n-\hat{\mathbf{u}}_{\mathtt{X}n}l|}&\approx&\frac{1}{R}, \\
\textrm{e}^{\textrm{i}k_0|\mathbf{r}-\mathbf{r}_n-\hat{\mathbf{u}}_{\mathtt{X}n}l|}&\approx&\textrm{e}^{\textrm{i}k_0(R-\mathbf{r}\cdot\hat{\mathbf{r}}_n)}, \nonumber \\
&& \\
\hat{\mathbf{u}}_{\mathtt{X}n}-\frac{(\mathbf{r}-\mathbf{r}_n-\hat{\mathbf{u}}_{\mathtt{X}n}l)}{|\mathbf{r}-\mathbf{r}_n-\hat{\mathbf{u}}_{\mathtt{X}n}l|}\frac{\hat{\mathbf{u}}_{\mathtt{X}n}\cdot(\mathbf{r}-\mathbf{r}_n-\hat{\mathbf{u}}_{\mathtt{X}n}l)}{|\mathbf{r}-\mathbf{r}_n-\hat{\mathbf{u}}_{\mathtt{X}n}l|}&\approx&\hat{\mathbf{u}}_{\mathtt{X}n} \\
\frac{(\mathbf{r}-\mathbf{r}_n-\hat{\mathbf{u}}_{\mathtt{X}n}l)\times\hat{\mathbf{u}}_{\mathtt{X}n}}{|\mathbf{r}-\mathbf{r}_n-\hat{\mathbf{u}}_{\mathtt{X}n}l|}&\approx&\hat{\mathbf{u}}_{\mathtt{X}n}\times\hat{\mathbf{r}}_n
\end{eqnarray}
and thus see (\ref{Ereduction1}) and (\ref{Breduction1}) reduce further still to
\begin{eqnarray}
\mathbf{E}&\approx&\Re\Bigg\{\sum_{\mathtt{X}=\mathtt{A},\mathtt{B}}\sum_{n=1}^N\frac{\textrm{i}\mu_0\omega_0I_{0\mathtt{X}n}W\textrm{e}^{\textrm{i}[k_0(R-\mathbf{r}\cdot\hat{\mathbf{r}}_n)-\omega_0 t+\varphi_{0\mathtt{X}n}]}}{4\pi R}\hat{\mathbf{u}}_{\mathtt{X}n}\Bigg\} \nonumber \\
&& \label{Ereduction2} \\
\mathbf{B}&\approx&\Re\Bigg\{\sum_{\mathtt{X}=\mathtt{A},\mathtt{B}}\sum_{n=1}^N\frac{\textrm{i}\mu_0\omega_0 I_{0\mathtt{X}n}W\textrm{e}^{\textrm{i}[k_0(R-\mathbf{r}\cdot\hat{\mathbf{r}}_n)-\omega_0 t+\varphi_{0\mathtt{X}n}]}}{4\pi cR}\hat{\mathbf{u}}_{\mathtt{X}n}\times\hat{\mathbf{r}}_n\Bigg\} \nonumber \\
&& \label{Breduction2}
\end{eqnarray}
with
\begin{eqnarray}
W&=&\int_{-L/2}^{L/2}w\textrm{d}l,
\end{eqnarray}
where we have used $L=\lambda_0/2\ll R$. Assuming that there are a large number of element pairs ($N \gtrsim 16\pi r^2_\textrm{max}/\lambda_0^2$) and that these are indeed uniformly distributed, we replace the discrete summation over $N$ with a continuous integral over $4\pi\,\textrm{sr}$ and thus see (\ref{Ereduction2}) and (\ref{Breduction2}) reduce finally to
\begin{eqnarray}
\mathbf{E}&\approx&\Re\Bigg\{\int_0^{2\pi}\!\!\!\int_0^\pi\frac{\textrm{i}\mu_0\omega_0 N W\textrm{e}^{\textrm{i}\{k_0[R-\mathbf{r}\cdot\hat{\mathbf{r}}(\vartheta,\varphi)]-\omega_0 t\}}}{16\pi^2R} \nonumber \\
&&\Big[I_{0\mathtt{A}}(\vartheta,\varphi)\textrm{e}^{\textrm{i}\varphi_{0\mathtt{A}}(\vartheta,\varphi)}\hat{\bm{\theta}}(\vartheta,\varphi) \nonumber \\
&+&I_{0\mathtt{B}}(\vartheta,\varphi)\textrm{e}^{\textrm{i}\varphi_{0\mathtt{B}}(\vartheta,\varphi)}\hat{\bm{\phi}}(\varphi)\Big]\textrm{sin}\vartheta\textrm{d}\vartheta\textrm{d}\varphi\Bigg\} \label{Ereduction3} \\
\mathbf{B}&\approx&\Re\Bigg\{\int_0^{2\pi}\!\!\!\int_0^\pi\frac{\textrm{i}\mu_0\omega_0 N W\textrm{e}^{\textrm{i}\{k_0[R-\mathbf{r}\cdot\hat{\mathbf{r}}(\vartheta,\varphi)]-\omega_0 t\}}}{16\pi^2 cR} \nonumber \\
&&\Big[I_{0\mathtt{A}}(\vartheta,\varphi)\textrm{e}^{\textrm{i}\varphi_{0\mathtt{A}}(\vartheta,\varphi)}\hat{\bm{\phi}}(\varphi) \nonumber \\
&-&I_{0\mathtt{B}}(\vartheta,\varphi)\textrm{e}^{\textrm{i}\varphi_{0\mathtt{B}}(\vartheta,\varphi)}\hat{\bm{\theta}}(\vartheta,\varphi)\Big]\textrm{sin}\vartheta\textrm{d}\vartheta\textrm{d}\varphi\Bigg\}, \label{Breduction3}
\end{eqnarray}
where we have introduced continuous peak electric current functions $I_{0\mathtt{X}}=I_{0\mathtt{X}}(\theta,\phi)$ as well as continuous phase constant functions $\varphi_{0\mathtt{X}}=\varphi_{0\mathtt{X}}(\theta,\phi)$ defined such that $I_{0\mathtt{X}n}=I_{0\mathtt{X}}(\hat{\mathbf{r}}_n)$ and $\varphi_{0\mathtt{X}n}=\varphi_{0\mathtt{X}}(\hat{\mathbf{r}}_n)$ ($\mathtt{X}\in\{\mathtt{A},\mathtt{B}\}$, $n\in\{1,\dots,N\}$). To generate an (approximate) unusual electromagnetic disturbance of the first kind we take the $I_{0\mathtt{X}}$ and the $\varphi_{0\mathtt{X}}$ to be constants, corresponding to each $\mathtt{A}$-type element having the same (possibly zero) peak electric current and phase constant and similarly for each $\mathtt{B}$-type element; (\ref{Ereduction3}) and (\ref{Breduction3}) become
\begin{eqnarray}
\mathbf{E}&\approx&\Re\Bigg\{E_0\Big[\tilde{\mathtt{A}}'_0\left(\hat{\mathbf{s}}g+\hat{\mathbf{z}}h\right)+\textrm{i}\tilde{\mathtt{B}}'_0\hat{\bm{\phi}}f\Big]\textrm{e}^{-\textrm{i}\omega_0t}\Bigg\} \label{Ereduction4} \\
\mathbf{B}&\approx&\Re\Bigg\{\frac{E_0}{c}\Big[\textrm{i}\tilde{\mathtt{A}}'_0\hat{\pmb{\phi}}f-\tilde{\mathtt{B}}'_0\left(\hat{\mathbf{s}}g+\hat{\mathbf{z}}h\right)\Big]\textrm{e}^{-\textrm{i}\omega_0t}\Bigg\} \label{Breduction4}
\end{eqnarray}
with
\begin{eqnarray}
E_0\tilde{\mathtt{A}}'_0&=&\frac{\textrm{i}\mu_0 \omega_0 NW}{8\pi R}I_{0\mathtt{A}}\textrm{e}^{\textrm{i}(k_0R+\varphi_{0\mathtt{A}})} \label{Aprimed} \\
E_0\tilde{\mathtt{B}}'_0&=&-\frac{\textrm{i}\mu_0 \omega_0 NW}{8\pi R}I_{0\mathtt{B}}\textrm{e}^{\textrm{i}(k_0R+\varphi_{0\mathtt{B}})}. \label{Bprimed}
\end{eqnarray}
Comparing (\ref{Ereduction4}) and (\ref{Breduction4}) with (\ref{EUED}) and (\ref{BUED}) we see that the antenna can indeed generate an (approximate) unusual electromagnetic disturbance of the first kind, as desired. A rotated and/or translated version of this disturbance can be generated using suitably modified forms for the $I_{0\mathtt{X}}$ and/or the $\varphi_{0\mathtt{X}}$. To generate an (approximate) unusual electromagnetic disturbance of the second kind we need only superpose the electric current distributions associated with each of its constituent unusual electromagnetic disturbances of the first kind.

In summary, our antenna is a spherical array of radiating dipolar elements configured to approximate the desired electromagnetic field near its centre. This design is arguably the simplest possible, given the reciprocal-space interpretation of the unusual electromagnetic disturbances (``... a continuous spherical superposition of plane electromagnetic waves...'').

%
%

\section{\label{A specific embodiment}A specific embodiment}
In this section we consider a specific embodiment of our antenna configured to generate an approximation of an electromagnetic tangle with $\tilde{\mathtt{A}}_0'=1/\sqrt{2}$ and $\tilde{\mathtt{B}}_0'=\textrm{i}/\sqrt{2}$ at a frequency of $f_0=2.45\,\textrm{GHz}$. We work in vacuum and focus on the electric field.

\begin{figure}
\includegraphics[width=\columnwidth]{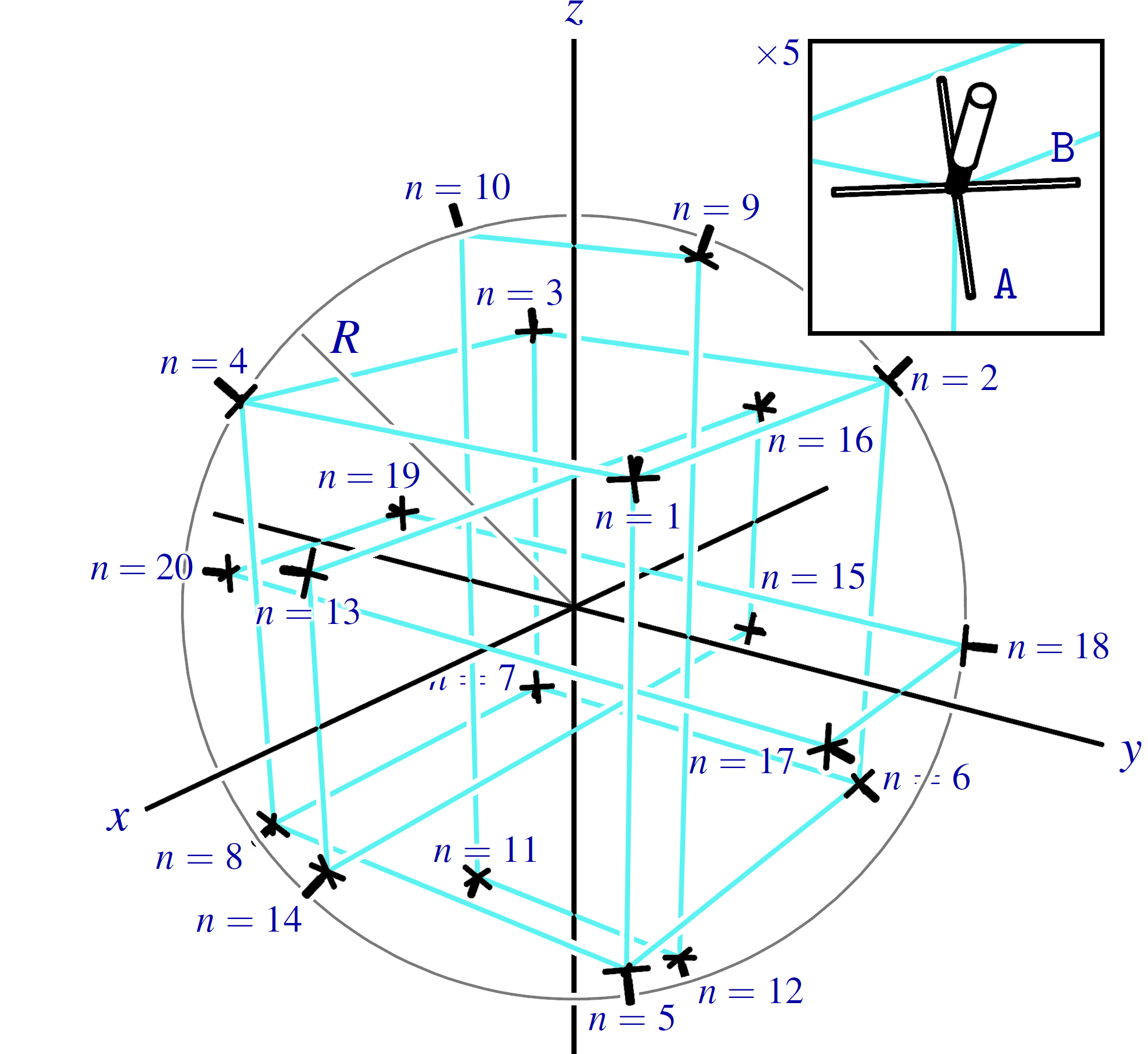}
\caption{\label{Figure1} The specific embodiment of our antenna considered in section \ref{A specific embodiment}. Included is a $5\times$ magnified view of the $n=1$ element pair.}
\end{figure}

Our embodiment consists of $2N=40$ elements in the form of half-wave dipoles ($L=\lambda_0/2=6.12\,\textrm{cm}$), arranged in pairs on the vertices of a dodecahedron \cite{Footnote3} with circumradius equal to five wavelengths ($R=5\lambda_0=61.2\,\textrm{cm}$) as indicated in Table~\ref{Table1} and shown in Fig.~\ref{Figure1}. We take
\begin{eqnarray}
I_{0\mathtt{A}}\textrm{e}^{\textrm{i}\varphi_{0\mathtt{A}}}&=&I_0\textrm{e}^{\textrm{i}(-k_0R-\pi/2)} \label{chosenA} \\
I_{0\mathtt{B}}\textrm{e}^{\textrm{i}\varphi_{0\mathtt{A}}}&=&I_0\textrm{e}^{\textrm{i}(-k_0R+\pi)}, \label{chosenB}
\end{eqnarray}
where $I_0$ is a peak electric current. This corresponds to $\tilde{\mathtt{A}}_0'\propto 1/\sqrt{2}$ and $\tilde{\mathtt{B}}_0'\propto\textrm{i}/\sqrt{2}$, as can be seen by comparing (\ref{chosenA}) and (\ref{chosenB}) with (\ref{Aprimed}) and (\ref{Bprimed}).

We only expect our antenna to generate a \textit{good} approximation of the exact electromagnetic tangle within a spherical region of radius
\begin{eqnarray}
r_\textrm{max}&=&\lambda_0\sqrt{\frac{N}{16\pi}} \nonumber \\
&=&0.63\lambda_0 \label{maxradius}
\end{eqnarray}
centred on the origin ($\mathbf{r}=0$). This conclusion can be reached by considering the surface area of the sphere to be divisible into $N$ patches of area equal to a half wavelength squared ($\lambda^2_0/4)$; within the spherical region the (discrete) array of elements that comprise the antenna is essentially indistinguishable from the continuous distribution required in the exact case. Outwith the spherical region the discrete nature of the array becomes apparent and we expect the approximation to break down accordingly.

\begin{figure}
\includegraphics[width=\columnwidth]{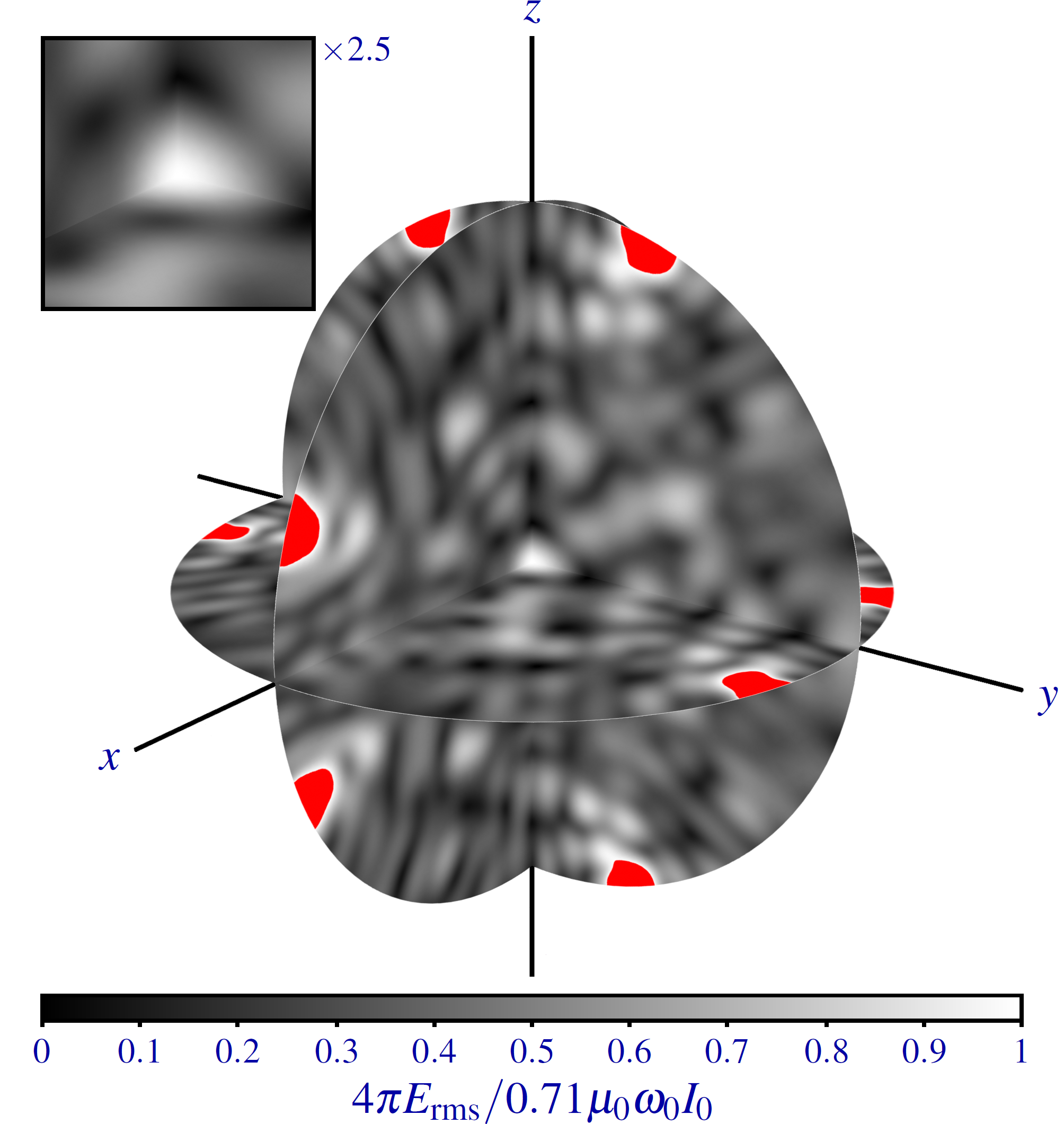}
\caption{\label{Figure2} A 3D composite of 2D density plots of the root-mean-square $E_\textrm{rms}$ of the electric field in the $x=0$, $y=0$ and $z=0$ planes inside the specific embodiment of our antenna considered in section \ref{A specific embodiment} ($0\le|\mathbf{r}|\le R$). Included is a $2.5\times$ magnified view centred on the origin ($\mathbf{r}=0$). Regions where $4\pi E_\textrm{rms}/0.71\mu_0\omega_0I_0>1$ are coloured red; these coincide with the positions of element pairs. The density plots were calculated numerically.}
\end{figure}

To quantify the degree of localisation of our (approximate) electromagnetic tangle we numerically calculated the root-mean-square $E_\textrm{rms}=E_\textrm{rms}(\mathbf{r})$ of the electric field, given by
\begin{eqnarray}
E_\textrm{rms}&=&\sqrt{\frac{1}{T_0}\int^{t_0+T_0}_{t_0}|\mathbf{E}|^2\textrm{d}t},
\end{eqnarray}
where $t_0$ is an arbitrary initial time. The variation of $E_\textrm{rms}$ is shown in Fig.~\ref{Figure2}, where it can be seen that $E_\textrm{rms}$ takes on a local maximum value of $0.71\mu_0 \omega_0 I_0/4\pi$ at the origin ($\mathbf{r}=0$); the (approximate) electromagnetic tangle is reasonably well localised, as desired. 

\begin{figure*}
\includegraphics[width=\textwidth]{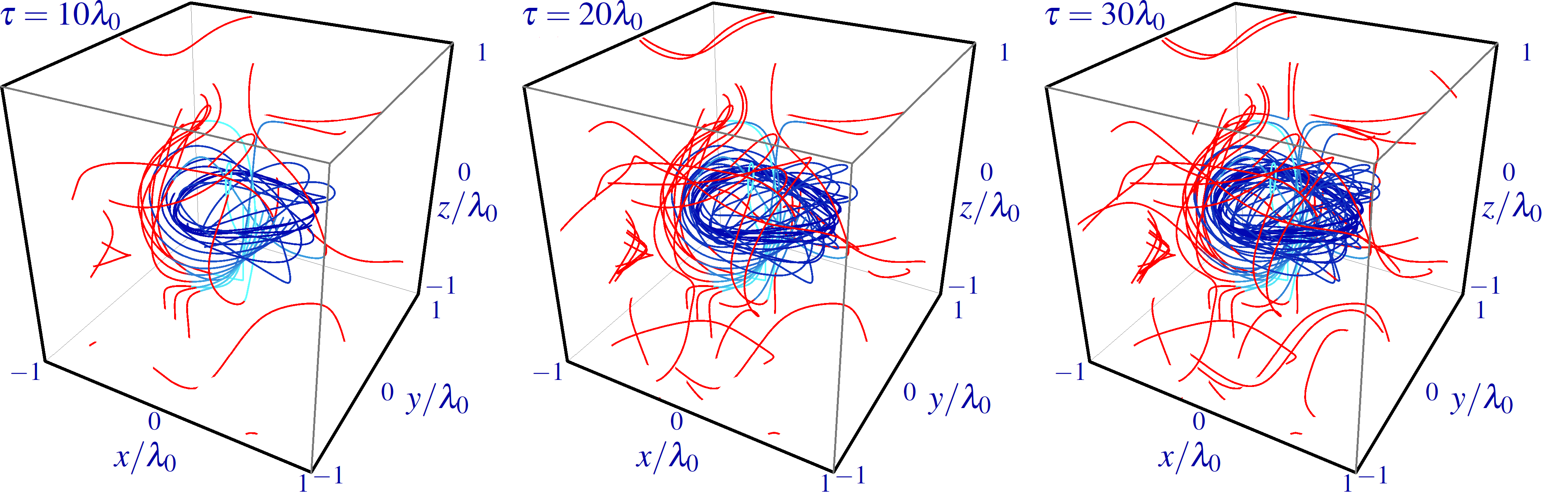}
\caption{\label{Figure3} Streamline plots of electric field lines inside the specific embodiment of our antenna considered in section \ref{A specific embodiment}. A total of $11$ electric field lines have been seeded at $11$ positions equally spaced along the positive $x$ axis in the range $(0, \lambda_0/2]$. Segments of electric field lines that lie outwith a spherical region of radius $r_\textrm{max}=0.63\lambda_0$ centred on the origin ($\mathbf{r}=0$) are coloured red; these segments are not expected to behave in the desired manner. The streamlines were calculated numerically using the fourth-order Runge-Kutta method with a step size equal to $\lambda_0/100$.}
\end{figure*}

\begin{figure*}
\includegraphics[width=\textwidth]{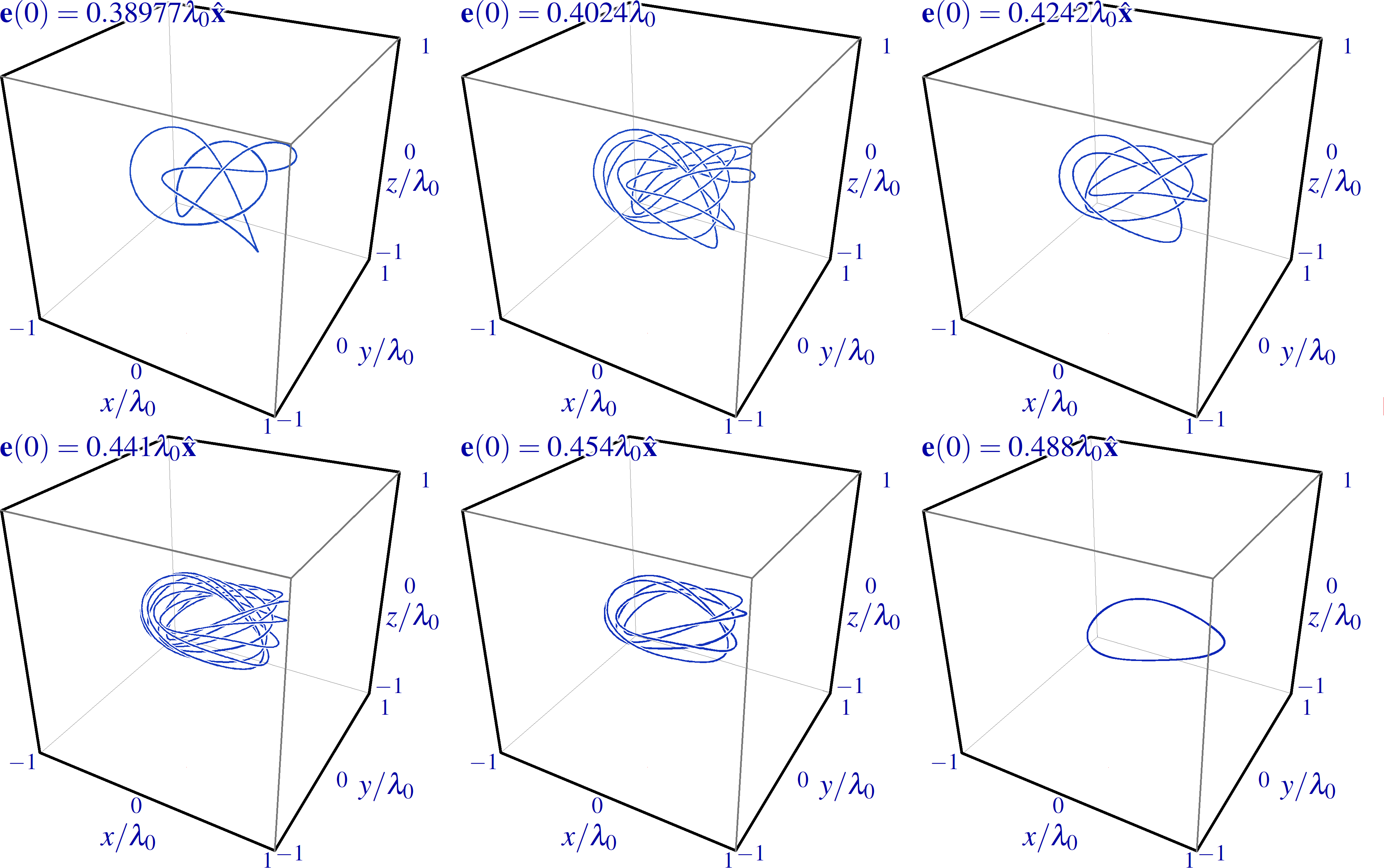}
\caption{\label{Figure4} Streamline plots of $5$ individual torus-knotted electric field lines and a trivially `knotted' electric field line inside the specific embodiment of our antenna considered in section \ref{A specific embodiment}. The seed position $\mathbf{e}(0)$ of each field line has been chosen carefully such that the line closes on itself within the integration length. The streamlines were calculated numerically using the fourth-order Runge-Kutta method with a step size equal to $\lambda_0/100$.}
\end{figure*}

To quantify the structure of the electric field lines of our (approximate) electromagnetic tangle at time $t$, we numerically integrated the streamline equation
\begin{eqnarray}
\frac{\textrm{d}\mathbf{e}(\tau)}{\textrm{d}\tau}&=&\frac{\mathbf{E}[\mathbf{e}(\tau),t]}{|\mathbf{E}[\mathbf{e}(\tau),t]|},
\end{eqnarray}
giving the trajectory $\mathbf{e}=\mathbf{e}(\tau)$ of an electric field line as
\begin{eqnarray}
\mathbf{e}(\tau)&=&\mathbf{e}(0)+\int_0^\tau\frac{\mathbf{E}[\mathbf{e}(\tau'),t]}{|\mathbf{E}[\mathbf{e}(\tau'),t]|}\textrm{d}\tau',
\end{eqnarray}
where $\mathbf{e}(0)$ is the seed position, $\tau$ is the integration length and we have assumed that the electric field is non-zero ($\mathbf{E}[\mathbf{e}(\tau'),t]\ne 0$, $0\le \tau'\le \tau$). The exact electromagnetic tangle is an electromagnetic standing wave with cylindrical symmetry about the $z$ axis and an electric field that vanishes everywhere at time $t=(1/2+ q)T_0/2$, where $q\in\{0,\pm1,\dots\}$ is an integer; the structure of the electric field lines at other times ($t\ne(1/2+ q)T_0/2$) is static and each electric field line passes through the positive $x$ axis. For the sake of brevity and without significant loss of generality we focus, therefore, on the structure of the electric field lines of our (approximate) electromagnetic tangle at $t=2\pi q$ with seed positions on the positive $x$ axis ($\mathbf{e}(0)=e_x(0)\hat{\mathbf{x}}$ with $0<e_x(0)<r_\textrm{max}$). A collection of electric field lines with different seed positions are shown in Fig.~\ref{Figure3} for integration lengths of $\tau=10\lambda_0$, $\tau=20\lambda_0$ and $\tau=30\lambda_0$, where a dichotomy can be seen: electric field lines seeded close to the $z$ axis quickly exit the spherical region $0\le|\mathbf{r}|\lesssim r_\textrm{max}$ and go on to trace out chaotic paths quite distinct from those seen in the exact electromagnetic tangle, in accord with the discussion surrounding (\ref{maxradius}); electric field lines that remain within the spherical region $0\le|\mathbf{r}|\lesssim r_\textrm{max}$, however, are torus knots (including the trivial `knot') qualitatively similar to those found in the exact electromagnetic tangle, as desired. A selection of the latter electric field lines are plotted individually in Fig.~\ref{Figure4} for $\tau=30\lambda_0$, where their torus-knotted forms can be seen clearly. Note that these electric field lines co-exist and are linked with each other.

We note for completeness that our antenna also generates linked and torus-knotted magnetic field lines with an overall structure similar to that of the electric field lines only shifted in time by a quarter cycle, as is the case for the exact electromagnetic tangle.

In summary, a specific embodiment of our antenna with a radius of $R=5\lambda_0=61.2\,\textrm{cm}$ and only $N=20$ element pairs driven at a frequency of $f_0=2.45\,\textrm{GHz}$ can yield linked and torus-knotted electric and magnetic field lines approximating those of an exact electromagnetic tangle, as desired.

\begin{table}
\caption{\label{Table1} Cartesian coordinates of the element pairs for the specific embodiment of our antenna considered in section \ref{A specific embodiment}, where $g=(1+\sqrt{5})/2$ is the golden ratio.}
\begin{ruledtabular}
\begin{tabular}{cccc|cccc}
$n$ & $\sqrt{3}x_n/R$ & $\sqrt{3}y_n/R$ & $\sqrt{3}z_n/R$ & $n$ & $\sqrt{3}x_n/R$ & $\sqrt{3}y_n/R$ & $\sqrt{3}z_n/R$ \\
\hline
$1$&$1$&$1$&$1$ & $11$& $0$&$-1/g$&$-g$ \\                  
$2$&$-1$&$1$&$1$ & $12$& $0$&$1/g$&$-g$ \\
$3$&$-1$&$-1$&$1$ & $13$& $g$&$0$&$1/g$ \\
$4$&$1$&$-1$&$1$ & $14$& $g$&$0$&$-1/g$ \\
$5$&$1$&$1$&$-1$ & $15$& $-g$&$0$&$-1/g$ \\
$6$&$-1$&$1$&$-1$ & $16$& $-g$&$0$&$1/g$ \\
$7$&$-1$&$-1$&$-1$ & $17$& $1/g$&$g$&$0$ \\
$8$&$1$&$-1$&$-1$ & $18$& $-1/g$&$g$&$0$ \\
$9$&$0$&$1/g$&$g$ & $19$& $-1/g$&$-g$&$0$ \\
$10$&$0$&$-1/g$&$g$ & $20$& $1/g$&$-g$&$0$ \\
\end{tabular}
\end{ruledtabular}
\end{table}

%
%

\section{\label{Outlook}Outlook}
There is much still to be done, most notably the experimental realisation of our antenna.

We have taken our antenna to be embedded in a simple medium; a transparent, isotropic, homogeneous, linear dielectric. It might prove fruitful to consider other, more exotic media such as microwave metamaterials, for example. We thank an anonymous referee for this suggestion.

Our antenna could be used to locally excite plasma. One motivation for doing so is the possible connection hypothesised in \cite{Boerner19a} between unusual electromagnetic disturbances and the as-yet unexplained natural phenomenon of ball lightning \cite{Shmatov19a}; a modern take, perhaps, on some old ideas \cite{Ohtsuki91a, Ranada96a}.

Although our focus in this paper has been on the microwave domain, we recognise that analogous ideas might be pursued in other frequency domains. For example: a monchromatic electromagnetic knot might be generated in the visible domain by superposing multiple laser beams or perhaps more elegantly by exciting the appropriate modes in an optical cavity.

%
%

\section*{Acknowledgements}
This work was supported by the Royal Society (URF$\backslash$R1$\backslash$191243), NWO/OCW (Quantum Software Consortium) and the Julian Schwinger Foundation (JSF-16-04-0000). RPC is a Royal Society University Research Fellow.

%
%

%
%


\begin{thebibliography}{99}
\bibitem{Footnote1} Strictly speaking, the ``electromagnetic cloud'' described in \cite{Cameron18a} is not a standing wave.
\bibitem{Footnote2} The element pairs should be located on the vertices of a tetrahedron for $N=4$, an octahedron for $N=6$, a cube for $N=8$, an icosahedron for $N=12$ or an icosahedron for $N=20$, these being the Platonic solids. Appropriate locations must be determined numerically for other values of $N$.
\bibitem{Footnote3} We investigated simpler designs with $N=4$, $N=6$, $N=8$ and $N=12$ but found that these could not generate a satisfactory approximation of an exact electromagnetic tangle.
\end{thebibliography}
\end{document}